\documentclass{article}[11pt]
\usepackage[dvips]{graphics,color}
\usepackage{graphicx}

\begin{document}

\begin{center}
{\bf \large SOLUTION OF THE RADIATIVE TRANSFER EQUATION IN THE SEPARABLE APPROXIMATION}
\vspace{0.7cm}

{\large G. V. Efimov}
\vspace{0.4cm}

{\small Bogoliubov Laboratory of Theoretical Physics, \\
Joint Institute for Nuclear Research, RU 141980, Dubna, Russia\\
(e-mail: efimovg@thsun1.jinr.ru) }
\vspace{0.4cm}

{\large N. V. Kryzhevoi}
\vspace*{0.4cm}

{\small Institut f\"ur Theoretische Astrophysik,\\
Universit\"at Heidelberg, Tiergartenstr. 15, D 69121 Heidelberg, Germany \\
Interdisziplin\"ares Zentrum f\"ur Wissenschaftliches Rechnen, \\
Universit\"at Heidelberg,
Im Neuenheimer Feld 368, D 69120 Heidelberg, Germany \\
(e-mail: kolya@ita.uni-heidelberg.de)}
\vspace*{0.4cm}

{\large  W. von Waldenfels}
\vspace*{0.4cm}

{\small Institut f\"ur Angewandte Mathematik,\\
Universit\"at Heidelberg, Im Neuenheimer Feld 294, D 69120 Heidelberg, Germany}
\vspace*{0.4cm}

{\small and}
\vspace*{0.4cm}

{\large   R. Wehrse}
\vspace*{0.4cm}

{\small Institut f\"ur Theoretische Astrophysik,\\
Universit\"at Heidelberg, Tiergartenstr. 15, D 69121 Heidelberg, Germany \\
Interdisziplin\"ares Zentrum f\"ur Wissenschaftliches Rechnen, \\
Universit\"at Heidelberg,
Im Neuenheimer Feld 368, D 69120 Heidelberg, Germany \\
(e-mail: wehrse@ita.uni-heidelberg.de)}
\end{center}

\begin{abstract}
A method for the fast and accurate solution of the radiative transfer equation 
in plane-parallel media with coherent isotropic scattering is presented.
This largely analytical approach uses the formalism of meromorphic functions
in order to obtain the total solution, i.e. the angle and depth variation of the
radiation field.  A discretization of space and angle coordinates is not required.
For the further application in the accretion disks a particular case of a 
finite slab whose the midplane is the symmetry plane is considered. 
\end{abstract}

\section{Introduction}
The transport of energy by means of radiation plays a very important role in, 
e.g., plasma physics, environmental physics,  and especially in astrophysics. 
The quantity most commonly used to describe a radiation field is the 
specific intensity, which obeys the radiative transfer equation (RTE) 
(cf.\cite{oxen}). In most cases this is an integro-differential equation. In 
many applications (e.g. stars) it is sufficient to consider a 1D radiative transfer
equation. The plane-parallel RTE can also be useful in modeling of accretion disks.
In general, an accretion disk is a complicated 3D structure and should be 
treated as such, however the multi-dimensional RTE remains computationally
costly. An alternative method used by various authors \cite{mayo, elkhoury, shaviv}
is to represent an accretion disk as a system of independent rings, each of 
them radiating as a plane-parallel slab.  Although such an assumption is unrealistic,
it nevertheless represents a landmark test in the theory of accretion disks. 

The 1D radiative transfer equation is rather well developed and there exists 
a vast literature on this topic ( see e.g. \cite{chandra, mihalas, cannon, 
kalk1, kalk2}). Among numerical methods the discrete ordinate method is the
most usable. The approximation of the definite integral in the RTE by a
quadrature sum leads to the replacement of the continuous radiation field by
a finite set of pencils of radiation. The subsequent discretization of the
transport operator results in a system of 
linear equations which is usually solved iteratively by Jacobi iteration 
("accelerated $\Lambda$-iteration"). This accurate and fast method fails, 
however, in  media with large optical depths and/or a large scattering fraction. 
Problems may also appear in media with steep gradients. 

Methods which proceed analytically as far as possible are most desirable. Usually
they give more insight into the general behavior of the solution and are less
CPU-time and memory consuming.  A method developed by Chandrasekhar 
\cite{chandra} can be included in this class of methods. It was found that the 
exact solution of the RTE in semi-infinite atmospheres leads to closed expressions
for the angular distribution of the emergent radiation, involving a so-called 
$H$-function that is the solution of an integral equation of standard form. In 
media of finite optical depth the emergent radiation can be expressed in terms 
of certain rational functions $X$ and $Y$ which satisfy integral equations too. 
The method is not commonly used because the Chandrasekhar functions $H$, $X$ and 
$Y$ are difficult to calculate in spite of the different methods proposed 
\cite{caldwell, bosma, haggag1, haggag2}. In addition, it does not provide the 
distribution of the specific intensity with depth that is required in various 
models of astrophysical objects. 

In the present paper we continue discussing a method for the analytical solution
of the plane-parallel RTE in finite media. In the previous papers (\cite{eww1, eww2})
the attention was mostly paid to mathematical aspects of the problem and 
algorithmic aspects were hardly considered. Only the angular distribution of the 
emergent intensity was found and only in the lowest "separable" approximation 
(see below). In this paper we solve an extended problem:
we also look for the solution of the RTE at every point
in a slab. We have slightly altered initial conditions and now consider a medium
with no incident radiation but with a non-zero depth-dependent source function.
The midplane of the medium is the symmetry plane which implies a further 
application of the solution to accretion disks. 

By introducing intensities in outward and inward directions we are able to write 
the RTE in a matrix form. The involved matrix $M$ is an operator in an infinite
dimensional space and possesses rather unpleasant properties, in particular,
its spectrum extends from $-\infty$ to $\infty$. In spite of this, the
formal solution can be expressed in terms of hyperbolic functions of the matrix
$M$. Note that these functions 
are bounded and therefore no difficulties arise related to the infinite
spectrum of $M$ . The successive application of the formalism of meromorphic
functions and Krein's formula for the finding of inverse operators
leads to the exact solution  represented in a form of infinite series. Because
of their very slow convergence these results are not well suited for
numerical work. Therefore an appropriate approximation of the infinite sums by 
finite ones is proposed where the number of terms defines the order of the
separable approximation. Experience has shown that the final results obtained
in such a matter in the lowest approximation are quite reasonable for  
practical purposes. In the 5--6th approximation orders the results can be 
regarded as highly precise.

\section{Formulation of the problem}
The specific intensity ${\mathcal I}(\tau,\mu)$ in  plane-parallel media with 
coherent isotropic scattering is described by the following radiative transfer
equation (cf. \cite{oxen})
\begin{equation}
\label{rte1}
\mu\frac{d {\mathcal I}(\tau,\mu)}{d\tau}=-{\mathcal I}(\tau,\mu)+\beta
\int\nolimits_{-1}^{+1}{\mathcal I}(\tau,\mu')\, d\mu' +\varepsilon B(\tau)
\end{equation}
where $B(\tau)$ is the Planck function, $\mu=\cos\theta$, $\theta$ is  
the angle between the normal and the beam of radiation, $\tau$ is optical depth, 
$\beta=(1-\varepsilon)/2$.
The de-excitation coefficient $\varepsilon$ can vary in the interval $[0,1]$.
We assume it is constant.

Our aim is to find the solution of Eq. (\ref{rte1}), i.e.  both the depth and 
angle variations of the radiation field in a slab of finite optical depth.  
We assume that the symmetry plane is at $\tau=0$. The optical depth 
is measured away from the symmetry plane and equals $\Delta$ and $-\Delta$ at 
the upper and lower boundaries, respectively.   

For the boundary conditions we require that no radiation is incident on the slab 
from outside, i.e. 
\begin{eqnarray}
\label{gru}
{\mathcal I}^+(\Delta, \mu)={\mathcal I}^-(-\Delta, \mu)=0
\end{eqnarray}
where
${\mathcal I}^+(\tau,\mu)$ and  ${\mathcal I}^-(\tau,\mu)$ are intensities 
in the positive and negative directions with respect to $\mu$, i.e. 
$$ {\mathcal I}^+(\tau,\mu)={\mathcal I}(\tau,\mu),~~~
{\mathcal I}^-(\tau,\mu)={\mathcal I}(\tau,-\mu)~~~
{\rm for}~~~\mu>0.$$

\section{Matrix form of the solution}
It is convenient to solve this problem in the matrix formalism.
The representation of the specific intensity as a two-component vector  
\begin{eqnarray*}
\!\!\!\!\!&& {\mathcal I}(\tau,\mu)=\left(\begin{array}{c}
   {\mathcal I}^+(\tau,\mu) \\
   {\mathcal I}^-(\tau,\mu) \end{array}\right)
 ={1\over\sqrt{\mu}}\,\left(\begin{array}{c}
   I^+(\tau,\mu) \\
   I^-(\tau,\mu) \end{array}\right)={1\over\sqrt{\mu}}\, {\bf I}(\tau,\mu)
\end{eqnarray*}
enables us to write Eq. (\ref{rte1}) in compact matrix notation
\begin{eqnarray}
\label{rte2}
\frac{d{\bf I}(\tau,\mu)}{d\tau} =- {\bf M}_{\mu\mu'}\,{\bf I}(\tau,\mu')+
{\bf B}(\tau)
\end{eqnarray}
where the integration over the index $\mu'$ is implied. The matrix ${\bf M}$
and the vector ${\bf B}$ are given by
\begin{eqnarray*}
{\bf M}_{\mu\mu'}={\bf D}_{\mu\mu'}-
(|{\bf v}\rangle\,\beta\, \langle{\bf u}|)_{\mu\mu'},~~~
{\bf B}(\tau)=\varepsilon B(\tau)\vert{\bf v}\rangle
\end{eqnarray*}
with
\begin{eqnarray*}
{\bf D}_{\mu\mu'}=\frac1\mu\,\tau_3\,\delta(\mu-\mu'),~ 
|{\bf v}\rangle=\frac1{\sqrt{\mu}}|e_-\rangle,~ 
|{\bf u}\rangle=\frac1{\sqrt{\mu}}|e_+\rangle,&& \\
\tau_3=
\left(\begin{array}{rr}
1&0 \\
0&-1
\end{array}\right),~~~~
|e_-\rangle=
\left(\begin{array}{r}
1 \\
-1
\end{array}\right),~~~~
|e_+\rangle=
\left(\begin{array}{r}
1 \\
1
\end{array}\right).&&
\end{eqnarray*}
The matrices ${\bf M}$ and $\tau_3$ do not commute but satisfy the condition 
\begin{eqnarray*}
\tau_3{\bf M}^{\top }={\bf M}\tau_3
\end{eqnarray*}

Subsequently we shall drop the index $\mu$ in the notation and  use
the following abbreviation
\begin{eqnarray*}
F({\bf M})|{\bf v}\rangle&=&
\int_0^1\frac{d\mu'}{\sqrt{\mu'}}\,
F({\bf M})_{\mu\mu'}|e_-\rangle,\\
\langle{\bf u}|F({\bf M})|{\bf v}\rangle&=&
\int\!\!\!\int_0^1\frac{d\mu d\mu'}{\sqrt{\mu \mu'}}\,
\langle e_+|F({\bf M})_{\mu\mu'}|e_-\rangle.
\end{eqnarray*}

The boundary conditions (\ref{gru}) can be represented in the following form
\begin{eqnarray*}
{\bf I}(\Delta)=\left(\begin{array}{c}
I_{\rm out} \\
0
\end{array}\right), &\qquad&
{\bf I}(-\Delta)=\left(\begin{array}{c}
0 \\
I_{\rm out}
\end{array}\right)
\end{eqnarray*}
or 
\begin{eqnarray}
\label{bc}
{\bf I}(\Delta)=I_{\rm out}\,\frac{1+\tau_3}{2}|e_+\rangle,&\quad&
{\bf I}(-\Delta)=I_{\rm out}\,\frac{1-\tau_3}{2}|e_+\rangle
\end{eqnarray}
where $I_{\rm out}=I_{\rm out}(\mu)$ is the outgoing intensity.
 
The formal solution of Eq. (\ref{rte2}) can be written in two equivalent forms
\begin{eqnarray}
\label{sol1}
{\bf I}(\tau)&=&e^{- {\bf M} (\tau+\Delta)}{\bf I}(-\Delta)+
\int\limits_{-\Delta}^{\tau}\!e^{- {\bf M}(\tau-\tau')}\,{\bf B}(\tau')\,d\tau'\\
&=&e^{- {\bf M}(\tau-\Delta)}{\bf I}(\Delta)-
\int\limits^{\Delta}_{\tau}\!e^{- {\bf M}(\tau-\tau')}\,{\bf B}(\tau')\,d\tau'\nonumber
\end{eqnarray}
which results in
\begin{eqnarray}
\label{idelt}
{\bf I}(\Delta)&=&e^{- 2{\bf M}\Delta}{\bf I}(-\Delta)+
\int\limits_{-\Delta}^{\Delta}e^{- {\bf M}(\Delta-\tau')}\,
{\bf B}(\tau')\,d\tau'.
\end{eqnarray}
Multiplying both sides of (\ref{idelt}) by $e^{{\bf M}\Delta}$, substituting  
(\ref{bc}) and using  $B(\tau)=B(-\tau)$ we obtain 
\begin{eqnarray}
\label{iout}
I_{\rm out}(\mu)=\langle e_+|\frac{\varepsilon}{\tau_3+\tanh({\bf M}\Delta)}
\!\int\limits_{0}^{\Delta}\!\frac{\cosh({\bf M}\tau')}{\cosh({\bf M}\Delta)}
\,B(\tau')|{\bf v}\rangle\,d\tau'.&& 
\end{eqnarray}
The internal distribution of the radiation field is given by the solution of 
Eq. (\ref{rte1}). It is sufficient to solve this equation for ${\mathcal I}^+$,
since due to the symmetry ${\mathcal I}^-$ follows immediately
\begin{eqnarray}
\label{sym}
I^+(-\tau)=I^-(\tau).
\end{eqnarray}
Thus for the given mean intensity and the boundary conditions we have
\begin{eqnarray}
\label{itot}
{\mathcal I}^+(\tau,\mu) =\int\limits_{-\Delta}^{\tau}\!\!e^{-\frac1\mu(\tau-\tau')}
\left[(1-\varepsilon)J(\tau')+\varepsilon B(\tau')\right]\,\frac{d\tau'}{\mu}
\end{eqnarray}
The mean intensity is given by
\begin{eqnarray*}
J(\tau)&=&\frac12\int\nolimits_0^1\!
\left({\mathcal I}^+(\tau,\mu')+{\mathcal I}^-(\tau,\mu')\right)d\mu'\\
&=&\frac12\int\nolimits_0^1
\!\frac1{\sqrt{\mu'}}\left(I^+(\tau,\mu')+I^-(\tau,\mu')\right)d\mu'
=\frac14\langle{\bf u}|\left(I^+(\tau)+I^-(\tau)\right)|e_+\rangle.
\end{eqnarray*}
Using the definition of ${\bf I}(\tau)$ and Eq. (\ref{sym}) we obtain
\begin{eqnarray*}
{\bf I}(\tau)+{\bf I}(-\tau)=\left(I^+(\tau)+I^-(\tau)\right)|e_+\rangle.
\end{eqnarray*}
So that the mean intensity  becomes
\begin{eqnarray}
\label{mint1}
J(\tau)&=&\frac14\langle{\bf u}|{\bf I}(\tau)+{\bf I}(-\tau)\rangle.
\end{eqnarray}
Using Eq. (\ref{sol1}) and the  relation
\begin{eqnarray*}
{\bf I}(\Delta)+{\bf I}(-\Delta)=I_{\rm out}|e_+\rangle
\end{eqnarray*}
we obtain
\begin{eqnarray}
\label{sint}
{\bf I}(\tau)&=&{e^{-{\bf M}\tau}\over2\cosh( {\bf M}\Delta)}I_{\rm out}|e_+\rangle\\
&+&\int\limits_{-\Delta}^\Delta \left\{\Theta(\tau-\tau')
{e^{-{\bf M}(-\Delta+\tau-\tau')}\over2\cosh({\bf M}\Delta)}
-\Theta(\tau'-\tau){e^{-{\bf M}(\Delta+\tau-\tau')}\over2\cosh({\bf M}\Delta)}
\right\}{\bf B}(\tau')\,d\tau'\nonumber
\end{eqnarray}
where the unit step function $\Theta(x)$ is given by
\begin{eqnarray*}
\Theta(x)=\left\{{0\mbox{  for  } x \leq 0, \atop 1\mbox{  for  }x>0.}\right.
\end{eqnarray*}
The substitution of Eqs. (\ref{sint}) and (\ref{iout}) into (\ref{mint1}) leads to
\begin{eqnarray}
\label{jtau}
J(\tau)&=&\frac{\varepsilon}{4}\int\limits_{-\Delta}^\Delta
\left(G_1(\tau,\tau')+G_2(\tau-\tau')\right)\,B(\tau')\,d\tau'
\end{eqnarray}
where
\begin{eqnarray*}
&&G_1(\tau,\tau')=\left\langle{\bf u}\left|
\frac{\cosh({\bf M}\tau)}{\cosh({\bf M}\Delta)}
\frac{1}{\tau_3+\tanh({\bf M}\Delta)}
\frac{\cosh({\bf M}\tau')}{\cosh({\bf M}\Delta)}\right|{\bf v}\right\rangle\\
&&G_2(\tau-\tau')=\left\langle{\bf u}\left|\frac{\sinh({\bf M}(\Delta-|\tau-\tau'|))}
{\cosh({\bf M}\Delta)}\right|{\bf v}\right\rangle
\end{eqnarray*}

\section{Evaluation of the matrix elements}
In order to evaluate (\ref{iout}) and (\ref{jtau}) we apply the formalism of 
meromorphic functions and Krein's formula.

By definition, a meromorphic function is a function that is analytic, except
for a set of poles $\xi_m$.  For a meromorphic function $F(z)$
decreasing for $|z|\rightarrow\infty$ the following representation is valid
\begin{eqnarray}
\label{mer}
F(z)=\sum_m \frac{F_m}{z-\xi_m}
\end{eqnarray}
where  
\begin{eqnarray*}
F_m=\lim_{z\to \xi_m}(z-\xi_m)\,F(z).
\end{eqnarray*}
In our case we have the meromorphic functions
\begin{eqnarray*}
\frac{\cosh({\bf M} a)}{\cosh({\bf M}\Delta)},\quad 
\frac{\sinh({\bf M} a)}{\cosh({\bf M}\Delta)}, \quad |a|\leq\Delta
\end{eqnarray*}
which in accordance with (\ref{mer}) can be represented by
\begin{eqnarray}
\label{fm1}
F({\bf M})&=&\sum_m \frac{F_m}{{\bf M}-\xi_m}=
\sum_{m=-\infty}^{\infty} \frac{F_m}{{\bf D}+i y_m-|{\bf v}
\rangle\,\beta\, \langle{\bf u}|}
\end{eqnarray}
where the constants $F_m$ depend on the particular form of the function 
$F({\bf M})$, and  the simple poles are 
\begin{eqnarray*}
\xi_m=-i y_m=-\frac{i\pi}{\Delta }\left(\frac12+m\right).
\end{eqnarray*}

In order to calculate an inverse operator such as that in (\ref{fm1})
we use Krein's formula which states that
for an operator S acting in an appropriate space $\mathcal L$ and being of the
following form
\begin{eqnarray*}
\label{krein1}
S=H-\sum_{i,j}^N|V_i\rangle c_{ij}\langle W_j|=H-|V\rangle\, c\,\langle W|
\end{eqnarray*}
the inverse operator $S^{-1}$ can be represented by
\begin{eqnarray}
\label{krein}
S^{-1}&=&{1\over H}+{1\over H}|V\rangle U\langle W|{1\over H}
\end{eqnarray}
where
\begin{eqnarray*}
U=(1-cT)^{-1}c,&&
T_{ij}=\left\langle W_i\left\vert{1\over H}\right\vert V_j\right\rangle.
\end{eqnarray*}
$H$ is an operator, $|V_i\rangle$ and $|W_i\rangle$ are vectors in the 
space $\mathcal L$, $c_{ij}$ is a number matrix.

The application of  Krein's formula gives
\begin{eqnarray}
\label{fm}
F({\bf M})&=&F({\bf D})
+\sum_m{1\over {\bf D}+i y_m}\left\vert{\bf v}\left\rangle{\beta F_m\over C(i y_m)}
\right\langle{\bf u}\right\vert{1\over {\bf D}+i y_m}
\end{eqnarray}
where
\begin{eqnarray}
\label{cim}
C(i y_m)&=&C_m=1
-\left\langle{\bf u}\left\vert{\beta\over {\bf D}+i y_m}
\right\vert{\bf v}\right\rangle\\
&=&1-2\beta\int\nolimits_0^1\frac{1}{1+\mu^2 y_m^2}\,d\mu=
1-2\beta\frac{\arctan(y_m)}{y_m}\nonumber
\end{eqnarray}
Using (\ref{fm}) and (\ref{cim}) we obtain
\begin{eqnarray*}
F({\bf M})|{\bf v}\rangle&=&\left.\left.\sum_m\frac{F_m}{C_m}
\frac1{{\bf D}+i y_m}\right|{\bf v}\right\rangle,\\
\langle{\bf u}|F({\bf M})&=&\sum_m\frac{F_m}{C_m}\left\langle{\bf u}\left|
\frac1{{\bf D}+i y_m},\right.\right.\\
\langle{\bf u}|F({\bf M})|{\bf v}\rangle&=&\frac1{\beta}
\sum_m{F_m}\left(\frac1{C_m}-1\right).
\end{eqnarray*}
In order to evaluate these expressions we apply the following representation 
\begin{eqnarray}
\frac1{C_m}=1-\frac{2t_0}{C_1}\frac1{t_0^2+y_m^2}+
\int\nolimits_0^1\frac{\rho(t)}{1+y_m^2t^2}\,d t
\end{eqnarray}
where 
\begin{eqnarray*}
\rho(t)=\frac{2\beta}{\left(1+\beta t
\ln\left({1-t\over 1+t}\right)\right)^2+
\left(\pi\beta t\right)^2},
\end{eqnarray*}
$t_0$ is a positive root of the equation
\begin{eqnarray*}
 C(t_0)=1+{\beta\over t_0}\ln\left(
{1-t_0\over1+t_0}\right)=0
\end{eqnarray*}
and
\begin{eqnarray*}
C_1=\left.\frac{d C(t)}{dt}\right|_{t=t_0}=
\frac{1-t_0^2-2\beta}{t_0(1-t_0^2)}.
\end{eqnarray*}
Since
\begin{eqnarray*}
\sum_m\frac{F_m}{\mu^2+y_m^2}&=&\frac1{2\mu}(F(\mu)-F(-\mu)),\\
\sum_m\frac{F_m}{\mu^2+y_m^2}\frac{1}{{\bf D}+ i y_m}
&=&\frac{F({\bf D})}{\mu^2-{\bf D}^2}
-\frac1{2\mu}\left(\frac{F(\mu)}{\mu-{\bf D}}+
\frac{F(-\mu)}{\mu+{\bf D}}\right),
\end{eqnarray*}
we obtain
\begin{eqnarray}
\label{sinh}
\left\langle{\bf u}\left|\frac{\sinh({\bf M}(\Delta-|\tau-\tau'|))}
{\cosh({\bf M}\Delta)}\right|{\bf v}\right\rangle&=&\Psi(|\tau-\tau'|),\nonumber\\
\label{cosheta}
\left.\left.\frac{\cosh({\bf M}\tau)}
{\cosh({\bf M}\Delta)}\right|{\bf v}\right\rangle&=&
\left.\left.\Phi(\tau,\mu)\right|{\bf v}\right\rangle,\\
\label{coshchi}
\left\langle{\bf u}\left|\frac{\cosh({\bf M}\tau)}
{\cosh({\bf M}\Delta)}\right.\right.&=&
\left\langle{\bf u}\left|\Phi(\tau,\mu)\right.\right.\nonumber
\end{eqnarray}
where
\begin{eqnarray}
\Psi(\tau)&=&-\frac{2}{\beta C_1}\frac{\sinh(t_0(\Delta-\tau))}
{\cosh(t_0\Delta)}
+\frac1{\beta}\int\nolimits_0^1\frac{\rho(t)}{t}\frac{\sinh((\Delta-\tau)/t)}
{\cosh(\Delta/t)}d t 
\end{eqnarray}
and
\begin{eqnarray}
\Phi(\tau,\mu)&=&\frac{\cosh(\tau/\mu)}{\cosh(\Delta/\mu)}
-\frac{2t_0}{C_1}
\left(\frac{\cosh(\tau/\mu)}{\cosh(\Delta/\mu)}-\frac{\cosh(t_0\tau)}
{\cosh(t_0\Delta)}\right)\frac{\mu^2}{t_0^2\mu^2-1}\nonumber \\
&+&\int\nolimits_0^1\left(\frac{\cosh(\tau/\mu)}{\cosh(\Delta/\mu)}
-\frac{\cosh(\tau/t)}{\cosh(\Delta/t)}\right)\frac{\mu^2\rho(t)}{\mu^2-t^2}
\,dt.
\end{eqnarray}

The last operator which has to be represented in a form appropriate for 
numerical work is $(1+\tau_3\tanh({\bf M}\Delta))^{-1}$. The formula (\ref{fm}) 
gives the following representation (see \cite{eww1,eww2})
\begin{eqnarray}
\label{1tt3}
1+\tau_3\tanh({\bf M}\Delta)&=&w^{(0)}+w^{(1)}P_-- w^{(2)}P_+\\
&=&(w^{(0)}+w^{(1)})P_-+(w^{(0)}-w^{(2)})P_+\nonumber
\end{eqnarray}
where
$P_{\pm}=\frac12\,|e_{\pm}\rangle\langle e_{\pm}|$
are the projection operators with the properties:
\begin{eqnarray*}
  P_++P_-=1,\quad P_\pm^2=P_\pm,\quad P_\pm P_\mp=0.
\end{eqnarray*}
The operators $w^{(j)}=w^{(j)}(\mu,\mu')$, $(j=0,1,2)$ are given by
\begin{eqnarray}
\label{w0}
w^{(0)}(\mu,\mu')&=&\left[1+\tanh\left({\Delta\over\mu}\right)\right]\delta(\mu-\mu'),\\
\label{w1}
w^{(1)}(\mu,\mu')
&=&{4\beta\over\Delta}\sum\limits_{m=0}^\infty
{1\over C_m}\cdot{\sqrt{\mu}\over 1+y_m^2\mu^2}\cdot
{\sqrt{\mu'}\over 1+y_m^2\mu'^2},\\
\label{w2}
w^{(2)}(\mu,\mu')
&=&{4\beta\over\Delta}\sum\limits_{m=0}^\infty
{y_m^2\over C_m}\cdot{\mu^{3/2}\over 1+y_m^2\mu^2}\cdot
{\mu'^{3/2}\over 1+y_m^2\mu'^2}.
\end{eqnarray}
The operators $w^{(1)}$ and $w^{(2)}$ are symmetric,
positive definite and have finite traces \cite{eww1,eww2}, i.e.
\begin{eqnarray*}
{\rm Tr}\,w^{(j)}=\int_0^1 w^{(j)}(\mu,\mu)\, d\mu<\infty,\quad (j=1,2).
\end{eqnarray*}
The inversion  of (\ref{1tt3}) is given by
\begin{eqnarray*}
\frac1{1+\tau_3\tanh({\bf M}\Delta)}=\frac1{w^{(0)}+w^{(1)}}P_-+
\frac1{w^{(0)}-w^{(2)}}P_+
\end{eqnarray*}
so that
\begin{eqnarray}
\label{tanht3}
&&\!\!\!\!\!\!\!\frac1{\tau_3+\tanh({\bf M}\Delta)}=
\left(\frac1{w^{(0)}-w^{(2)}}-\frac1{w^{(0)}}\right)P_+\tau_3\\
&&~~~~~~~~~~~~~~~~~~-\left(\frac1{w^{(0)}}-\frac1{w^{(0)}+w^{(1)}}\right)P_-\tau_3+
\frac1{w^{(0)}}\tau_3\nonumber
\end{eqnarray}

\section{Separable representation of the solution}
Krein's formula (\ref{krein}) can now be applied to the expressions
in brackets in (\ref{tanht3}). $w^{(0)}$ corresponds to $H$ and the terms
\begin{eqnarray*}
{\sqrt{\mu}\over1+y_m^2\mu^2}\cdot{\sqrt{\mu'}\over1+y_m^2\mu'^2},
~~~~{\rm and}~~~~
{\mu^{3/2}\over1+y_m^2\mu^2}\cdot{\mu'^{3/2}\over1+y_m^2\mu'^2}
\end{eqnarray*}
which present in the definitions (\ref{w1}) and (\ref{w2}) can be regarded
as the vectors $V_m$ and $W_m$.
Unfortunately, the infinite sums converge very slowly and one needs up to several
thousands terms in order to achieve the required accuracy. Consequently the 
dimension of the matrices involved in (\ref{krein}) becomes very large that makes
the application of  Krein's formula inefficient here. In order to 
accelerate computations we use the following approximations for $w^{(1)}$ 
and  $w^{(2)}$.

The operators $w^{(1)}$ and $w^{(2)}$ are represented by
\begin{eqnarray}
\label{w1e}
w^{(1)}(\mu,\mu')&=&\sqrt{\mu}\cdot
{E(\mu^2)-E(\mu'^2)\over\mu^2-\mu'^2}\cdot
\sqrt{\mu'},\\
\label{w2e}
w^{(2)}(\mu,\mu')&=&\mu^{3/2}\cdot
{{E(\mu^2)\over\mu^2}-{E(\mu'^2)\over\mu'^2}\over
\mu'^2-\mu^2}\cdot\mu'^{3/2}
\end{eqnarray}
where
\begin{eqnarray}
\label{fe}
E(\mu^2)&=&{4\beta\over\Delta}\sum\limits_{m=0}^\infty
{1\over C_m}\cdot{\mu^2\over 1+y_m^2\mu^2}.
\end{eqnarray}
Let us approximate the function $E(t)$ by a finite sum
\begin{eqnarray}
\label{fea}
E(t)\approx E_N(t)=\sum\limits_{n=1}^N
{a_nt\over 1+A_nt}.
\end{eqnarray}
Then  we  get for (\ref{w1e}) and (\ref{w2e})
\begin{eqnarray}
\label{4.6}
w^{(1)}(\mu,\mu')&\approx&\sum\limits_{nn'}^N
V_n^{(1)}(\mu)\, a_n\delta_{nn'}\, V_{n'}^{(1)}(\mu') 
=\vert V^{(1)}\rangle J^{(1)}\langle V^{(1)}\vert,\\
\label{4.7}
w^{(2)}(\mu,\mu')
&\approx&\sum\limits_{nn'}^NV_n^{(2)}(\mu)\,a_nA_n\delta_{nn'}\,
V_{n'}^{(2)}(\mu')
=\vert V^{(2)}\rangle J^{(2)}\langle V^{(2)}\vert
\end{eqnarray}
where
\begin{eqnarray*}
V_n^{(1)}(\mu)={\sqrt{\mu}\over 1+A_n\mu^2},&&
J_{nn'}^{(1)}=a_n\delta_{nn'},\\
V_n^{(2)}(\mu)={\mu^{3/2}\over1+A_n\mu^2},&&
J_{nn'}^{(2)}=a_nA_n\delta_{nn'}.
\end{eqnarray*}
The number $N$ in (\ref{fea}) defines the $N$-th separable 
approximation.

Taking into account the representations (\ref{4.6}) and  (\ref{4.7}) and using
Krein's formula (\ref{krein}) we get now
\begin{eqnarray}
&&\!\!\!\!{1\over w^{(0)}}-{1\over w^{(0)}+w^{(1)}}
=\left.{1\over w^{(0)}}\left\vert V^{(1)}\left\rangle
S^{(1)}\right\langle V^{(1)}
\right\vert{1\over w^{(0)}}\right.,\nonumber\\
&&\!\!\!\!{1\over w^{(0)}-w^{(2)}}-{1\over w^{(0)}}=
\left.{1\over w^{(0)}}\left\vert V^{(2)}\left\rangle
S^{(2)}\right\langle V^{(2)}
\right\vert{1\over w^{(0)}}\right.\nonumber
\end{eqnarray}
where
\begin{eqnarray*}
S^{(1)}_{nn'}&=&\left({1\over 
1+J^{(1)}U^{(1)}}\, J^{(1)}\right)_{nn'}, \ 
S^{(2)}_{nn'}=\left({1\over 
1-J^{(2)}U^{(2)}}\, J^{(2)}\right)_{nn'}
\end{eqnarray*}
and
\begin{eqnarray*}
U^{(1)}_{nn'}&=&{1\over2}\int\limits_0^1
{\mu\left(1+e^{-\frac{2\Delta}{\mu}}\right)\over
(1+A_n\mu^2)(1+A_{n'}\mu^2)}\,d\mu,\\
U^{(2)}_{nn'}&=&{1\over2}\int\limits_0^1
{\mu^3\left(1+e^{-\frac{2\Delta}{\mu}}\right)\over
(1+A_n\mu^2)(1+A_{n'}\mu^2)}\,d\mu.
\end{eqnarray*}
The substitution of the third term of (\ref{tanht3})  into (\ref{jtau}) gives
\begin{eqnarray*}
\left\langle{\bf u}\left|\frac{\cosh({\bf M}\tau')}
{\cosh({\bf M}\Delta)}\frac1{w^{(0)}}\tau_3\frac{\cosh({\bf M}\tau)}
{\cosh({\bf M}\Delta)}\right|{\bf v}\right\rangle
=\int\nolimits_0^1\Phi(\tau,\mu)\Phi(\tau',\mu)
(1+e^{-\frac{2\Delta}{\mu}})\frac{d\mu}{\mu}.&&\nonumber
\end{eqnarray*}
The contribution of the second term of (\ref{tanht3}) equals zero because
$P_-\tau_3|e_-\rangle=0$. The substitution of the first term leads to 
\begin{eqnarray*}
&&\left\langle{\bf u}\left|\frac{\cosh({\bf M}\tau)}{\cosh({\bf M}\Delta)}
\left(\frac1{w^{(0)}-w^{(2)}}-\frac1{w^{(0)}}\right)P_+\tau_3
\frac{\cosh({\bf M}\tau')}{\cosh({\bf M}\Delta)}\right|{\bf v}\right\rangle\\
&&~~~=\left\langle{\bf u}\left|\frac{\cosh({\bf M}\tau)}{\cosh({\bf M}\Delta)}
\right|{\bf Y}^+_n\right\rangle S_{nn'}^{(2)}\left\langle {\bf Y}^-_{n'}\left|
\frac{\cosh({\bf M}\tau')}{\cosh({\bf M}\Delta)}\right|{\bf v}\right\rangle
\end{eqnarray*}
where
\begin{eqnarray*}
|{\bf Y}^{\pm}_{n}\rangle=\frac{V^{(2)}_n}{w^{(0)}}|e_{\pm}\rangle
\end{eqnarray*}
and
\begin{eqnarray*}
\left\langle{\bf u}\left|\frac{\cosh({\bf M}\tau)}{\cosh({\bf M}\Delta)}
\right|{\bf Y}^+_n\right\rangle&=&\left\langle {\bf Y}^-_n\left|
\frac{\cosh({\bf M}\tau)}{\cosh({\bf M}\Delta)}\right|{\bf v}\right\rangle\\
&=&\int\nolimits_0^1\Phi(\tau,\mu)\left(1+e^{-\frac{2\Delta}{\mu}}\right)
\frac{\mu}{1+A_n\mu^2}d\mu=K_n(\tau).
\end{eqnarray*}
Taking into account the above expressions the mean intensity becomes
\begin{eqnarray}
\label{saj}
J(\tau)&=&\frac{\varepsilon}{4}
\int\nolimits_{-\Delta}^\Delta\!\Psi(|\tau-\tau'|)B(\tau')\,d\tau'
+\frac{\varepsilon}{4}\,K_n(\tau)\,S_{nn'}^{(2)}
\int\nolimits_0^\Delta \! K_{n'}(\tau')B(\tau')\,d\tau'\nonumber\\
&+&\frac{\varepsilon}{2}\int\nolimits_0^1 \frac1{\mu}
\Phi(\tau,\mu)\left(1+e^{-\frac{2\Delta}{\mu}}\right)
\int\nolimits_0^\Delta \!\!\Phi(\tau',\mu)B(\tau')\,d\tau'd\mu
\end{eqnarray}
and the emergent intensity takes the form
\begin{eqnarray}
\label{iout2}
{\mathcal I}_{\rm out}(\mu)&=&\varepsilon
\left(1+e^{-\frac{2\Delta}{\mu}}\right)\left(\frac1{\mu}\int\nolimits_0^\Delta
\!\!\Phi(\tau',\mu)B(\tau')\,d\tau'\right.\nonumber\\
&+&\left.\frac{\mu}{2(1+A_n\mu^2)}\,S_{nn'}^{(2)}
\int\nolimits_0^\Delta \! K_{n'}(\tau')B(\tau')\,d\tau'\right).
\end{eqnarray}
where the Einstein's convention is implied.

\section{Discussions}
The practical efficiency of the method depends largely on how well the lower
orders of the approximation  represent the exact solution. 
In order to obtain an accurate solution in the low approximation
orders a clever approximation of the function $E(t)$ is necessary.
The representation of the original function (\ref{fe}) by the finite sum 
(\ref{fea}) with the minimal number of terms and without loss of  
accuracy is the crucial point of our investigation.
\begin{figure}[t]
\begin{center}
\scalebox {0.66} {\includegraphics{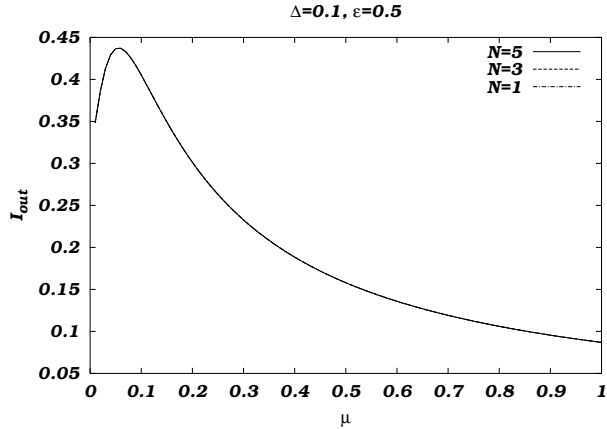}}
\caption{\rm Angular distribution of the outgoing intensity calculated in the 
different approximation orders by means of equation (\ref{itot}) with the Planck 
function $B(\tau)=1-0.5\left(\tau/\Delta\right)^2$. The curves representing
the results of the different approximation orders coincide.}
\label{f1}
\end{center}
\end{figure}
\begin{figure}[t]
\begin{center}
\scalebox {0.66} {\includegraphics{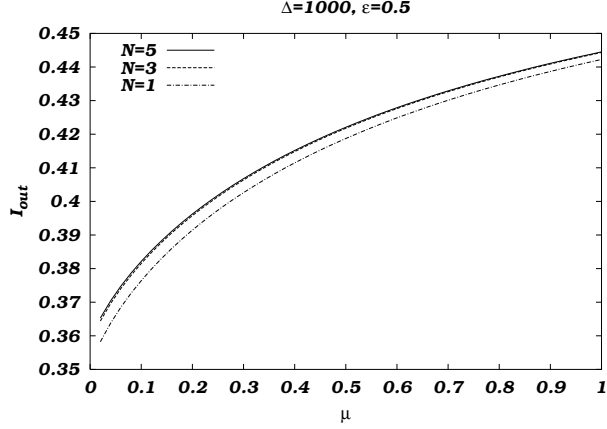}}
\caption{\rm Angular distribution of the outgoing intensity calculated in the 
different approximation orders by means of equation (\ref{itot}) with the Planck 
function $B(\tau)=1-0.5\left(\tau/\Delta\right)^2$.}
\label{f2}
\end{center}
\end{figure}

We propose two methods for the 
approximation (\ref{fea}). The first was originally developed by Stieltjes and 
Markov  (Appendix A). Using the theory of the orthonormal polynomials,
the coefficients $a_n$ and $A_n$ can be found very quickly. However, roundoff errors 
appearing in the calculation of the coefficients $p_{kl}$ in (\ref{pkl}) do not 
allow the method to be implemented, in particular, by means of FORTRAN codes. 
The method is well suited for programs like MATHEMATICA. The convergence of the
final solution is moderate in optically thick media and high in optically thin 
ones.

The second method is the so called "Points method". It consist of the solution 
of the system of $2N$ algebraic equations as described in Appendix B.
In contrast to the Stieltjes-Markov method it does not suffer from
roundoff errors, and FORTRAN codes run very well.  Some matrices
become badly conditioned in the high approximation orders ($N>6$). However these
cases imply a very high accuracy of the final solution, which is usually not needed
in applications, and are not considered.
In optically thin media the solution obtained with such $a_n$ and $A_n$ shows 
approximately the same convergence as in the Stieltjes-Markov method. 
The convergence of the "Points method" is much better in optically thick slabs. 
The calculation of $a_n$ and $A_n$ takes much longer because one needs the 
values of the original $E(t)$ in the reference points $\{t_1,...,t_{2N}\}$.  

 Further, all the calculations are carried out with the application of the 
"Points method" where the points are chosen as $t_i=i/2N$.

Figs. \ref{f1} and \ref{f2}  show the angular distribution of the emergent 
intensity in slabs of  different optical depth. The results were obtained 
through Eq. (\ref{itot}) with the mean intensity $J(\tau)$ calculated by means 
of (\ref{saj}). The dependence of the 
approximation order is shown. As one can see in optically thin media the 
precise results can be obtained already in the lowest separable approximation 
whereas in optically thick slabs a few additional approximation orders are 
necessary in order to match the exact solution. 

\begin{figure}[t]
\begin{center}
\scalebox {0.66} {\includegraphics{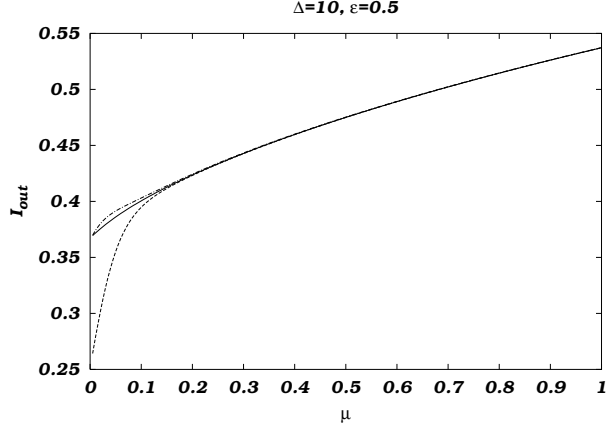}}
\end{center}
\caption{\rm Angular distribution of the outgoing intensity calculated in the
6-th approximation order with the Planck function 
$B(\tau)=1-0.5\left(\tau/\Delta\right)^2$. The solid curve represents the precise
results obtained by means of (\ref{itot}). Equation (\ref{iout2}) was used
to get the dotted curve.  The results obtained with the improved approximation
of the operator $w^{(2)}$ (\ref{w2n}) are represented by the dashed-dotted curve.} 
\label{fs3}
\end{figure}

However, the simplest and the fastest way for the calculation of the emergent
intensity is the direct application of (\ref{iout2}). As in the previous case
a few approximation orders are sufficient to obtain results with reasonable
accuracy. The exception is for the range of small $\mu$ (see Fig.~\ref{fs3}). 
The difference between this solution and the exact one is caused  by errors in 
the approximation of the function $E(\mu^2)$.
So, for $\mu  \rightarrow 0$ we have $E(\mu^2)\sim \mu$, whereas 
$E_N(\mu^2)\sim \mu^2$. Although the asymptotic behavior of 
$w^{(1)}(\mu,\mu')$ and $w_N^{(1)}(\mu,\mu')$ at small $\mu$ are the same,   
$w^{(2)}(\mu,\mu')$ and  $w_N^{(2)}(\mu,\mu')$  behave in different ways:
when $\mu,\mu' \rightarrow 0$ $w^{(2)}(\mu,\mu')\sim \sqrt{\mu}\sqrt{\mu'}$
while  $w^{(2)}_N(\mu,\mu')\sim \mu^{3/2}{\mu'}^{3/2}$.

Better results in the region of small $\mu$ can be obtained by means of a
further improvement of the approximation for the operator $w^{(2)}$: 
\begin{eqnarray}
\label{w2n}
w^{(2)}(\mu,\mu')&=&\mu^{3/2}\cdot
{{E(\mu^2)\over\mu^2}-{E(\mu'^2)\over\mu'^2}\over
\mu'^2-\mu^2}\cdot\mu'^{3/2}\nonumber\\
&=&\sqrt{\mu}\cdot
{{\mu'\over\mu}E(\mu^2)-{\mu\over\mu'}E(\mu'^2)\over
\mu'^2-\mu^2}\cdot\sqrt{\mu'}\nonumber\\
&=&w^{(1)}(\mu,\mu')+w^{(3)}(\mu,\mu')
\end{eqnarray}
where
\begin{eqnarray*}
w^{(3)}(\mu,\mu')&=&\sqrt{\mu}\cdot
{{1\over\mu}E(\mu^2)-{1\over\mu'}E(\mu'^2)\over
\mu'-\mu}\cdot\sqrt{\mu'}.
\end{eqnarray*}
Applying the "Points method" for the following approximation
\begin{eqnarray}
\label{fee}
{1\over\mu}E(\mu^2)&=&{4\beta\over\Delta}\sum\limits_{m=0}^\infty
{1\over C_m}\cdot{\mu\over 1+y_m^2\mu^2}\approx 
\sum\limits_{n=1}^N{b_n\over 1+B_n\mu}
\end{eqnarray}
we get 
\begin{eqnarray}
\label{w3}
w^{(3)}(\mu,\mu')&\approx&\sum\limits_{n=1}^N
{\sqrt{\mu}\over 1+B_n\mu}\cdot b_nB_n\cdot{\sqrt{\mu'}\over 1+B_n\mu'}.
\end{eqnarray}
Thus the final results calculated with the new representation of $w^{(2)}$ 
(\ref{w2n}) show  better agreement with the exact solution for all $\mu$
as shown in Fig.~\ref{fs3}.
\begin{figure}[t]
\scalebox {0.9} {\includegraphics{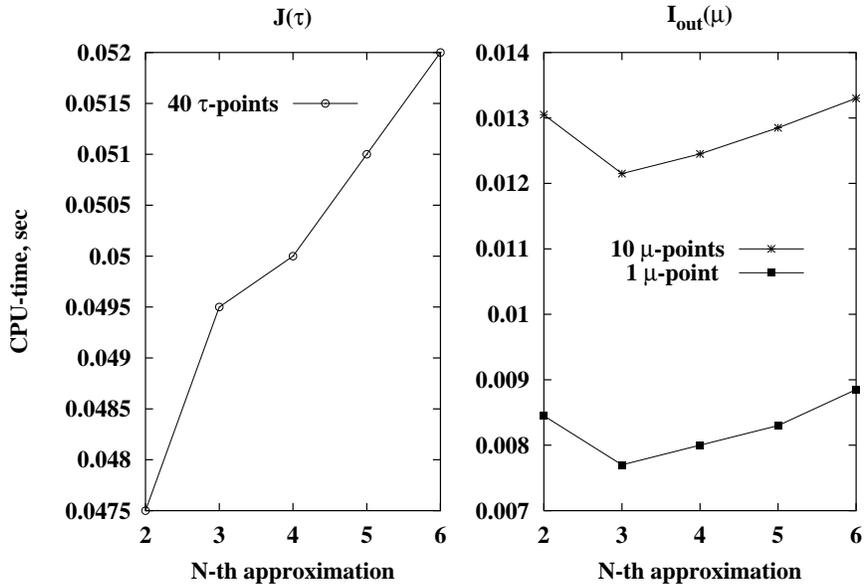}}
\caption{\rm CPU-time required for the calculation of the mean intensity
and the intensity emerging from the surface of a slab with $\Delta=100$
and $\varepsilon=0.5$ in the different approximation orders. }
\label{fs2}
\end{figure}

Fig. \ref{fs2} shows the CPU-time required for the calculation of the internal 
distribution of the mean intensity ({\it left panel}; Eq.~(\ref{saj})) and the 
angular distribution of the outgoing intensity ({\it right panel}; 
Eq.~(\ref{iout2})) in the different approximation orders. FORTRAN 
double-precision, optimized codes were used on a HP C240 computer.
 Note that time necessary for the 
calculations of the coefficients $a_n$, $A_n$ and matrix $S^{(2)}_{nn'}$ is
also included. Although these calculations are carried out only once for the
given $\Delta$ and $\varepsilon$, the most time (about 8 ms in the 2nd order)
is spent on this, which remains the place for further improvements.

The comparison of our results with those obtained by the finite element and 
finite difference methods in the case of isothermal media can be found in 
\cite{richl}.

\section{Summary}
In this paper we have given a new method for the solution of the plane-parallel
radiative transfer equation in media of finite optical depth.
The continuous angle and depth variation of the specific intensity including the
angular distribution of the emergent radiation can be obtained fast and accurately
for a wide range of  parameters.  The considered configuration of the
medium makes the accretion disk the most appropriate object for further
application of this solution. 

Generalization to the cases of anisotropic and non-coherent scattering is also 
possible. In particular the modeling of a spectral line with complete 
redistribution will be presented in a separate paper.

\vspace*{0.4cm}

We would like to thank K. Manson for valuable comments which have lead 
to an improvement of this paper.
This work was supported by the Deutsche Forschungsgemeinschaft 
(Sonderforschungsbereich 359/C2)
and the Graduier\-tenkolleg at the Interdisciplinary Center for Scientific 
Computing of Heidelberg University.

\appendix
\section{Approximation of $E(t)$: Stieltjes-Markov method}
\setcounter{equation}{0}
\def\theequation{\thesection.\arabic{equation}}
In order to represent $E(t)$  in the form of a finite sum 
\begin{eqnarray}
\label{feap}
E(t)\approx E_N(t)=\sum\limits_{n=1}^N{a_nt\over 1+A_nt}
\end{eqnarray}
one can apply the Stieltjes-Markov method (see \cite{perron}). 

Let us introduce a measure $\phi(dx)$ on interval $[0,1]$ in such a way that
\begin{eqnarray}
\label{meas}
\int\limits_0^1\phi(dx)f(x)={2\over\Delta}\sum\limits_{m=0}^\infty
{1\over 1+y_m^2}\cdot f\left({1\over 1+y_m^2}\right).
\end{eqnarray}
For the approximate evaluation of the integral we use a quadrature formula
\begin{eqnarray}
\label{quad}
\int\limits_0^1\phi(dx)f(x)\simeq\sum\limits_{n=1}^Nc_nf(x_n)
\end{eqnarray}
where $x_n$ are zeros of polynomials of degree $N$ which are orthogonal
with respect to $\phi(dx)$, i.e.
\begin{eqnarray*}
\int\limits_0^1\phi(dx)~P_m(x)P_l(x)=0~~~~~{\rm for}~~~~m\neq l,
\end{eqnarray*}
and the weights $c_n$ are given by  
\begin{eqnarray*}
c_n={Q_N(x_n)\over P_N'(x_n)}
\end{eqnarray*}
where $Q_k(x)$ are associated  polynomials of degree $(N-1)$ 
\begin{eqnarray*}
Q_k(s)=\int\limits_0^1\phi(dx){P_k(s)-P_k(x)\over s-x}.
\end{eqnarray*}
The polynomials $P_k(x)$ can be constructed in the 
following way. Let us calculate $2N+1$ moments
\begin{eqnarray}
m_k(\Delta)&=&\int\limits_0^1\phi(dx)x^k={2\over\Delta}\sum\limits_{m=0}^\infty
\left({1\over 1+y_m^2}\right)^{1+k}\nonumber\\
&=&\left.{1\over k!}\left(-{\partial\over\partial v}\right)^k
{\tanh\left({\Delta\sqrt{v}}\right)\over\sqrt{v}}\right\vert_{v=1}
\end{eqnarray}
A matrix $H$ with elements
\begin{eqnarray*}
H_{ij}=m_{i+j},~~~~~(i,j=0,...,N)
\end{eqnarray*}
is symmetric and positive definite. 
The Cholesky decomposition gives an upper-triangular matrix $r$ with the 
property that $H$ can be written as 
$ H=r^\top r$. Then the elements
\begin{eqnarray}
\label{pkl}
p_{kl}=\left(r^{-1}\right)^\top_{kl}
\end{eqnarray}
define the coefficients of the  polynomial $P_k(x)$
\begin{eqnarray*}
P_k(x)=\sum\limits_{l=0}^kp_{kl}x^l.
\end{eqnarray*}
The associated polynomial can be written as
\begin{eqnarray*}
Q_k(x)=\sum\limits_{l=0}^{k-1} x^l\sum\limits_{i=l+1}^k p_{ki}\,m_{i-l-1}. 
\end{eqnarray*}
Let us introduce the notation
\begin{eqnarray*}
&& t={s\over1+s},~~~~s={t\over1-t},~~~~0\leq s\leq\infty,\\
&& x={1\over1+y^2},~~~~0\leq x\leq1,~~~~y=\sqrt{{1-x\over x}}.
\end{eqnarray*}
Then we have 
\begin{eqnarray*}
E(t)={4\beta\over\Delta}\sum\limits_{m=0}^\infty
{1\over C(i y_m)}\cdot{t\over 1+y_m^2t}
={4\beta\over\Delta}\sum\limits_{m=0}^\infty
{1\over 1+y_m^2}\cdot {1\over C(i y_m)}\frac{s}{{1\over 1+y_m^2}+s}.&&
\end{eqnarray*}
According to (\ref{meas}) and (\ref{quad}) we get 
\begin{eqnarray*}
E(t)&=&\int\limits_0^1\phi(dx)
{\beta\over C\left(i\sqrt{{1-x\over x}}\right)}{s\over x+s}
\simeq\sum\limits_{n=1}^N
{c_n\beta\over C\left(i\sqrt{{1-x_n\over x_n}}\right)}{s\over s+x_n}\\
&=&\sum\limits_{n=1}^N{c_n\beta\over C\left(i\sqrt{{1-x_n\over x_n}}\right)}
{t\over x_n+(1-x_n)t}.
\end{eqnarray*}
Thus the coefficients in (\ref{feap}) become
\begin{eqnarray}
a_n={c_n\over x_n}\cdot{2\beta\over C\left(i\sqrt{A_n}\right)},
~~~~~~~A_n={1\over x_n}-1.
\end{eqnarray}

\section{Approximation of $E(t)$: "Points method"}
\setcounter{equation}{0}
\def\theequation{\thesection.\arabic{equation}}
The coefficient $a_n$ and $A_n$ of the $N$-th separable approximation
can be found by solving the system of algebraic equations: 
\begin{eqnarray}
\label{etl1}
E_N(t_l)&=&
\sum\limits_{n=1}^N{a_nt_l\over1+A_nt_l},~~~(l=1,...,2N)
\end{eqnarray}
where $\{t_1,...,t_{2N}\}$ are points in the interval $[0,1]$. 
The solution of (\ref{etl1}) can be obtained in the following way. We have
\begin{eqnarray}
\label{etl2}
{E(t_l)\over t_l}\prod\limits_{k=1}^N(1+A_kt_l)
&=&\sum\limits_{n=1}^Na_n\prod\limits_{k\neq n}(1+A_kt_l).
\end{eqnarray}
Let us introduce the notation
\begin{eqnarray*}
&& \prod\limits_{k=1}^N(1+A_kt_l)=
\sum\limits_{s=0}^Nt^s_lu_s,\\
&& u_0=1,~~~u_s=\sum\limits_{1\leq i_1<...<i_s\leq N}
A_{i_1}\cdot A_{i_2}\cdot...\cdot A_{i_s},\\
&& \prod\limits_{k\neq n}^{N}(1+A_kt_l)=
\sum\limits_{s=0}^Nt^s_lu_s\vert_{A_n=0}=
\sum\limits_{s=1}^Nt^{s-1}_lv_s^{(n)},\\
&& \sum\limits_{n=1}^Na_n\prod\limits_{k\neq n}(1+A_kt_l)=
\sum\limits_{s=1}^Nt_l^{s-1}b_s,\quad
b_s=\sum\limits_{n=1}^Na_nv_s^{(n)}.
\end{eqnarray*}
Then we can write (\ref{etl2}) as the following 
\begin{eqnarray}
\label{sle}
\sum\limits_{s=1}^Nt_l^{s-1}b_s-
\sum\limits_{s=1}^Nt_l^{s-1}E(t_l)u_s&=&{E(t_l)\over t_l}.
\end{eqnarray}
Let us introduce $N\times N$- matrices and $N$-component vectors
\begin{eqnarray*}
&& Q^{(1)}=Q^{(1)}_{ls}=t^{s-1}_l,~~~~~
 H^{(1)}=H^{(1)}_{ls}=t^{s-1}_lE(t_l),\\
&& Q^{(2)}=Q^{(2)}_{ls}=t^{s-1}_{N+l},~~~~~
 H^{(2)}=H^{(2)}_{ls}=t^{s-1}_{N+l}E(t_{N+l}),\\
&& f^{(1)}=f^{(1)}_l={E(t_l)\over t_l},~~~~~
f^{(2)}=f^{(2)}_l={E(t_{N+l})\over t_{N+l}}.
\end{eqnarray*}
So that the system of linear equations (\ref{sle}) can be rewritten in a form
\begin{eqnarray}
\label{ms}
&& Q^{(1)}b-H^{(1)}u=f^{(1)},\\
&& Q^{(2)}b-H^{(2)}u=f^{(2)}.\nonumber
\end{eqnarray}
Eliminating $b$ we obtain the vector $u=(u_1,...,u_N)$
\begin{eqnarray*}
u&=&\left[
(Q^{(1)})^{-1}H^{(1)}-(Q^{(2)})^{-1}H^{(2)}\right]^{-1}\\
&\times&\left[(Q^{(2)})^{-1}f^{(2)}-(Q^{(1)})^{-1}f^{(1)}\right].
\end{eqnarray*}
The roots of the following polynomial
\begin{eqnarray}
(-x)^N+u_1(-x)^{N-1}+...+u_N
=(-1)^N\prod\limits_{n=1}^N(x-A_n)=0&&\nonumber
\end{eqnarray}
correspond to the constants $A_1,....,A_N$. The last step is the solution
of the matrix equation
\begin{eqnarray*}
C\,a=f^{(1)}
\end{eqnarray*}
with
\begin{eqnarray*}
C_{ls}={1\over1+A_st_l},~~~~~(l,s=1,...,N)
\end{eqnarray*}
in order to get the coefficients $a=(a_1,...,a_N)$.


\begin{thebibliography}{}
\bibitem{oxen}
J. Oxenius: {\it Kinetic Theory of Particles and Photons}, Springer-Verlag, 
Berlin, 1986
\bibitem {mayo}
S. K. Mayo, D. T. Wickramasinghe and J. A. J. Whelan: 
{\it Mon.\ Not.\ R.\ astr.\ Soc.}~ {\bf 193}~ (1980), 793-824.   
\bibitem {shaviv}
G. Shaviv and R. Wehrse: 
{\it Astron.\ Astrophys.}~ {\bf 251}~ (1991), 117-132. 
\bibitem {elkhoury}
W. El-Khoury and D. T. Wickramasinghe : 
{\it Astron.\ Astrophys.}~ {\bf 358}~ (2000), 154-168. 
\bibitem {chandra} 
S. Chandrasekhar: 
{\it Radiative Transfer}, Dover Publications, New York, 1950.
\bibitem{mihalas}
D. Mihalas: 
{\it Stellar Atmospheres}, first edition, Freeman, San Francisco, 1970.
\bibitem{cannon} 
C. J. Cannon: 
{\it The Transfer of Spectral Line Radiation}, Cambridge University Press,
Cambridge, 1985.
\bibitem{kalk1} 
W. Kalkofen: 
{\it Methods in Radiative Transfer}, Cambridge University Press, Cambridge, 1984.
\bibitem{kalk2} 
W. Kalkofen: 
{\it Numerical Radiative Transfer}, Cambridge University Press, Cambridge, 1987.
\bibitem{caldwell}
J. Caldwell and A. J. Perks: 
{\it Astrophys.\ J.}~ {\bf 249}~ (1981), 258-262. 
\bibitem{bosma}
P. B. Bosma and W. A. de Rooij: 
{\it Astron.\ Astrophys.}~ {\bf 126}~ (1983), 283-292. 
\bibitem{haggag1}
M. H. Haggag and H. M. Machali: 
{\it Astrophys.\ Space\ Sci.}~ {\bf 111}~ (1985), 189-195. 
\bibitem{haggag2}
M. H. Haggag, H. M. Machali, S. El-Labany and M. T. Attia:
{\it Astrophys.\ Space\ Sci.}~ {\bf 149}~ (1989), 13-18. 
\bibitem{eww1}
G. V. Efimov, W. von Waldenfels and R. Wehrse: 
{\it J.\ Quant.\ Spectrosc.\ Radiat.\ Transfer}~ {\bf 53}~ (1995), 59-74.
\bibitem{eww2}
G. V. Efimov, W. von Waldenfels and R. Wehrse: 
{\it J.\ Quant.\ Spectrosc.\ Radiat.\ Transfer}~ {\bf 58}~ (1997), 355-373.
\bibitem{richl}
S. Richling, E. Meink\"ohn, N. Kryzhevoi and G. Kanschat: 
{\it Astron.\ Astrophys.}~ 2001, in press.
\bibitem {perron} 
O. Perron:  
{\it Die Lehre von den Kettenbr\"{u}chen}, Band II, B. G. Teubner 
Verlagsgesellschaft, Stuttgart, 1957.
\end{thebibliography}
\end{document}